\documentclass[usenatbib]{mn2e}
\usepackage{times}
\usepackage{epsfig}
\usepackage{amsmath}
\usepackage{amssymb}
\title[The accretion of galaxies into groups and clusters ]{The
accretion of galaxies into groups and clusters}

\author[McGee et al.]{Sean L. McGee$^{1}$\thanks{Email:
s2mcgee@uwaterloo.ca}, Michael L. Balogh$^{1}$, Richard G. Bower$^{2}$,
Andreea S. Font$^{2}$,\newauthor Ian G. McCarthy$^{3,4,5}$\\
$^{1}$Department of Physics and Astronomy, University of Waterloo,
Waterloo, Ontario, N2L 3G1, Canada\\ $^{2}$Department of Physics,
University of Durham, Durham, UK, DH1 3LE\\ $^{3}$Kavli Institute for
Cosmology, University of Cambridge, Cambridge, UK, CB3 0HA\\ 
 $^{4}$Astrophysics Group, Cavendish Laboratory, University of Cambridge,
Cambridge, UK, CB3 0HE\\ 
 $^{5}$Institute of Astronomy, University of Cambridge, Cambridge, UK, CB3
OHA\\ 
}

\date{\today}

\def\Mtrunc{$M_{\mathrm{trunc}}$} 
\def\Ttrunc{$T_{\mathrm{trunc}}$}
\def\LCDM{$\Lambda$CDM}
\def\Mdoth{$h^{-1}~$M$_\odot$}
\def\Mpch{$h^{-1}~$Mpc}
\begin{document}
\maketitle

\begin{abstract}  
 We use the galaxy stellar mass and halo merger tree information from
 the semi-analytic model galaxy catalogue of \citet{font} to examine
 the accretion of galaxies into a large sample of groups and clusters,
 covering a wide range in halo mass (10$^{12.9}$ to 10$^{15.3}$
 \Mdoth), and selected from each of four redshift epochs (z=0, 0.5,
 1.0 and 1.5).  We find that clusters at all examined redshifts have
 accreted a significant fraction of their final galaxy populations
 through galaxy groups.  A 10$^{14.5}$ \Mdoth\ mass cluster at z=0 has,
 on average, accreted $\sim$ 40$\%$ of its galaxies (M$_{stellar} >
 10^{9}$ \Mdoth ) from halos with masses greater than 10$^{13}$
 \Mdoth. Further, the galaxies which are accreted through groups are
 more massive, on average, than galaxies accreted through smaller
 halos or from the field population.  We find that at a given epoch,
 the fraction of galaxies accreted from isolated environments is
 independent of the final cluster or group mass. In contrast, we find
 that observing a cluster of the same halo mass at each redshift epoch
 implies different accretion rates of isolated galaxies, from 5-6 $\%$
 per Gyr at z=0 to 15$\%$ per Gyr at z=1.5. We find that combining the
 existence of a Butcher Oemler effect at z=0.5 and the observations
 that galaxies within groups display significant environmental effects
 with galaxy accretion histories justifies striking
 conclusions. Namely, that the dominant environmental process must
 begin to occur in halos of 10$^{12}$ -- 10$^{13}$ \Mdoth, and act over
 timescales of $>$ 2 Gyrs. This argues in favor of a
 mechanism like ``strangulation'', in which the hot halo of a galaxy
 is stripped upon infalling into a more massive halo . This simple
 model predicts that by z=1.5 galaxy groups and clusters will display
 little to no environmental effects. This conclusion may limit the
 effectiveness of red sequence cluster finding methods at high
 redshift.

\end{abstract}


\section{Introduction}

In recent years, an extraordinary confluence of independent
measurements of the cosmological parameters has resulted in the
concordance model of the Universe (\LCDM), in which the mass density
is dominated by cold dark matter. In this model, the initial
distribution of density perturbations has the greatest power on small
scales, which causes low mass dark matter haloes to form first at high
redshift. Larger haloes form later through the merging, or accretion,
of smaller halos. Eventually, this 'hierarchical structure formation'
leads to the formation of galaxy groups and clusters, which become
more common with time. The mass assembly of dark matter halos has been
extensively studied analytically \citep{press_schechter, bond_eps,
bower_eps, lacey_cole, sheth_tormen, vandenbosch, benson_merger} and
through numerical simulations \citep{DEFW, limogao}. Consistent with
these studies, \citet[][hereafter B09]{berrier} used n-body
simulations to show that the mass assembly of clusters is dominated by
the most massive accretion events; in effect, the merging of groups
with clusters. However, by associating dark matter subhalos with
galaxies, they show that the {\it galaxy} assembly of clusters is
dominated by lower mass halos, or the infalling of isolated
galaxies. This distinction could be of great importance since there
are a variety of physical processes that depend on the mass of the
host dark matter halo and which could affect the properties of a
galaxy, such as ram pressure stripping, strangulation and galaxy
harassment.

Indeed, detailed observations of dense environments, galaxy groups,
and clusters in the local universe have shown that the galaxies which
inhabit these environments have properties substantially different
from galaxies in low density or field environments. In particular,
galaxy groups and clusters have lower average galaxy star formation
rates \citep{lewis, gomez}, lower fractions of disk galaxies
\citep{dressler,mcgee}, and higher red fractions
\citep{balogh_bimodal,weinmann} than field galaxies. Despite this
wealth of observational data, there is no consensus on the dominant
physical mechanism responsible for these differences, mainly because
large populations of ``transition'' objects have avoided detection. In
particular, there is no large excess in the fraction of galaxies
between the red sequence and the blue cloud in dense environments
\citep{balogh_bimodal, weinmann}. While there are specific examples of
transitioning spiral galaxies which are in the process of having their
HI gas stripped due to ram pressure in local clusters
\citep{KenneyHI,VollmerHI}, the X-ray temperatures and pressures, as
well as the infalling velocity of the galaxies, required for such a
transformation mechanism are probably too high to be effective in low
mass groups.

Strangulation, the process in which the more loosely bound hot halo of
a galaxy is stripped by the group or cluster halo, leaving a reduced
amount of gas available for future star formation
\citep{balogh_model}, is an attractive candidate because it is still
effective in low mass groups \citep{mccarthy, kawata}. However, it is
not clear if such a gentle mechanism can account for the dramatic
effect seen in clusters. \citet{zabludoff} have proposed that the
extreme properties of galaxy clusters may result from the
``pre-processing'' of galaxies in group environments before accretion
into the cluster. This is supported by observations of reduced star
formation rates in the outskirts of clusters, well past the virial
radius \citep{balogh_cnoc1,lewis}. However, B09 have
claimed that ``pre-processing'' is not a large effect. They find only
$\sim$ 12 $\%$ of galaxies are accreted in to the final cluster
environment as members of groups with five or more galaxies. While the
B09 clusters are relatively low mass, their work
shows the importance of distinguishing the accretion of galaxies from
that of dark matter mass.

A complementary approach to trying to isolate ``transition galaxies''
is to study the properties of galaxies in groups and clusters as a
function of redshift. As first shown by \citet{ButcherOemler1} and
confirmed by many others \citep[eg.][]{lavery,couch,ellingson}, the
fraction of blue galaxies in clusters increases with redshift, the so
called Butcher-Oemler effect. Despite this, the fraction of star
forming galaxies in groups and clusters is still lower than the coeval
field fraction at least to z=1
\citep{wilmancnoc,gerke,Balogh_cnoccol}.  The need to explain the
Butcher-Oemler effect, as well as the local properties of galaxy
clusters provides important constraints for the nature of the
transformation mechanism. Essentially, if the transformation mechanism
only occurs in very massive clusters, then the fraction of blue
galaxies is simply the fraction of galaxies which have fallen into the
cluster within the time scale of transformation.

The time scale for transformation of galaxy properties
to occur is a significant uncertainty in attempting to link the growth
of structure to the Butcher-Oemler effect. Previous attempts using
cluster assembly histories adopted relatively short time scales of
$\sim$ 1 Gyr and, while complicated by uncertain cosmological
parameters, showed that a direct infall model alone did not produce
enough evolution in the blue fraction \citep{bower_eps,Kauffmann}.
\citet{kodamabower} combined the evolving star formation properties of
field galaxies with a cluster infall model to successfully reproduced
the scatter in the red sequence of low redshift clusters.  Similarly,
\citet{ellingson} found that the radial distribution of early type
galaxies in galaxy clusters at two redshift epochs could best be
explained if the galaxy infall into clusters decreased by a factor of
$\sim$ 3 between z $>$ 0.8 and z $\sim$ 0.5.

In this paper, we examine the ${\it galaxy}$ assembly properties of
groups and clusters over a wide mass range and at four redshift
epochs.  We investigate the mass of halos through which groups and
clusters gain their galaxies and the extent to which preprocessing in
the group environment is important at four redshift epochs. By making
simple assumptions, we investigate the predictions for the fraction of
galaxies in groups and clusters which are ``environmentally affected''
for a range of relevant timescales and the halo mass thresholds which
those effects begin. Using these models we try to gain insight into
the dominant physical processes necessary to reproduce observations of
group and cluster galaxies, as well make predictions for future
observations.  In \textsection \ref{sec-data}, we present the details
of our simulated clusters and some of their properties and present our
results in \textsection \ref{sec-results}. We discuss these results
and conclude in \textsection \ref{sec-conclude}. In this paper, we
assume a cosmology with $\Omega_m$ = 0.25, $\Omega_\Lambda$ = 0.75,
$\sigma_8$=0.9 and $H_0$ = 100$h$ km s$^{-1}$ Mpc$^{-1}$ unless
mentioned otherwise.

\section{Simulations} \label{sec-data}

To interpret observations of galaxy properties as a function of
environment, we need to know the accretion history of those galaxies;
as shown by B09 this can be subtly different from the total mass
assembly history.  But galaxy formation has proven to be a difficult
problem, and it is not clear, given that the dark matter halo mass
function has a very different shape from the galaxy luminosity
function, if the approach of B09 of simply associating subhalos with
galaxies includes all of the relevant physics. At the least, this
approach does not allow for the robust identification of the stellar
masses of galaxies. Unfortunately, an obvious alternative --- the
direct simulation of the baryonic processes of galaxies --- is
difficult on the scale of the cosmological volumes needed to study
large samples of groups and clusters.

Semi-analytic galaxy formation models provide a good tool to
encapsulate the essential physical processes of gas cooling, star
formation and feedback \citep[e.g.][]{whitefrenk, kauffmann93,
somerville, croton, bowermodel}. Dark matter simulations, on which
modern semi-analytic models are based, are now large enough to allow
the study of the growth of the groups and clusters over a wide range
of redshifts.  We make use of one such semi-analytic model by
\citet[][hereafter F08]{font}, which is a recent modification to the
Durham semi-analytic model (GALFORM) of \citet{bowermodel}. The basic
prescriptions for gas cooling and star formation in the GALFORM model
was laid out by \citet{Colemodel}, and subsequently modified for
modern cosmological parameters by \citet{Bensonmodel}. The model of
\citet{bowermodel} introduced a method for parameterizing the effect
of AGN feedback on the gas in massive galaxies to correct for the
``overcooling'' problem.

The \citet{bowermodel} model, as in essentially all previous
semi-analytic models, implements a relatively simple treatment of
environmental effects, in which the hot gas reservoirs of galaxies are
removed upon becoming a satellite galaxy. Many authors have since
shown that this approach produces an unphysically high fraction of red
galaxies in groups and clusters \citep{weinmann, baldry06,
gilbank}. The F08 model implements a more realistic ``strangulation''
model in which the hot gas halo of galaxies falling into more massive
halos are removed according to a prescription of
\citet{mccarthy}. However, a careful examination of cluster and group
data with this model at a range of redshifts reveal that there are
important discrepancies. In particular, the model overpopulates the
green valley between the blue cloud and red sequence \citet[McGee et
al., in prep.]{Balogh_cnoccol}. We emphasize that despite this
difficulty in reproducing galaxy colours, the stellar masses of
galaxies in the F08 and Bower et al. models are much better
understood. In particular, the Bower et al. model reproduces the
observed evolution of the stellar mass function out to at least z=5.

In this paper, our analysis relies primarily on the GALFORM prediction
of the stellar mass function of galaxies in different environments.
This is insensitive to the problem noted above, as the star formation
rate of galaxies declines rapidly with redshift, so the bulk of a
galaxy's stellar mass is already in place before it ever becomes a
satellite. Thus, the details of the strangulation procedure adopted in
GALFORM are unimportant for our analysis and, indeed, all our
conclusions are independent of the choice of either the Bower et
al. model or the F08. model.

\subsection{Cluster and group sample} \label{sec-sample}

The F08 model, from which our simulated galaxy clusters and groups are
drawn, is based on merger trees derived from the dark matter
Millennium simulation \citep{MillSim}, a $\Lambda$CDM cosmological box
with 500/$h$ Mpc sides. The Millennium simulation uses
\textsc{GADGET2} \citep{gadget2}, a TREE-PM N-body code, and an
initial power spectrum calculated using \textsc{CMBFAST}
\citep{cmbfast}. The merger trees are generated as described in
\citet{helly_trees} and \citet{harker_trees}, and are complete down to
halos which host $\sim$ 10$^{8}$ \Mdoth\ galaxies. In this paper, we
are principally concerned with selecting samples of galaxies which are
observationally accessible, and thus specify a single fixed stellar
mass cut of $M>10^{9}$ \Mdoth, much higher than the completeness
limit.

We analyze all the groups and clusters in the F08 model more
massive than $M$ = 10$^{12.9}$ \Mdoth\ at four redshift epochs (z=0,
0.5, 1 and 1.5). The key properties of the cluster samples are shown
in Table \ref{table-clusterprops}. In particular, we show the number
of clusters, and the average number of galaxies with stellar masses
above $M$ = 10$^{9}$ \Mdoth\ at the epoch of observation, in each of
the mass bins which will be used in the remainder of the paper.

\begin{figure}
\leavevmode \epsfysize=8cm \epsfbox{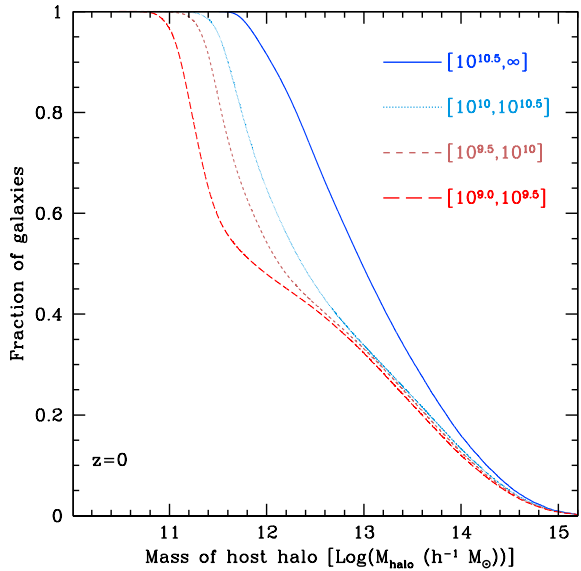}
\caption{The cumulative distribution of the host halo mass of galaxies
  at z=0. The distribution is shown for four ranges
  in the galaxy's stellar mass at $z=0$, shown in the upper right corner
  in units of \Mdoth.}
\label{residing}
\end{figure}

In Figure \ref{residing}, we present the cumulative distribution of
galaxies which reside within the virial radius of host halos of a
given mass. We plot this for four stellar mass ranges at z=0.  In the
F08 model, $\sim$ 50$\%$ of z = 0 L$_*$ galaxies are in host halos
with masses above 10$^{12.5}$ \Mdoth.  This compares very well with
observational results: \citet{berlind} found that $\sim$ 56 $\%$ of
M$_r <-20.5$ galaxies in the SDSS are linked to groups containing at
least one other member, a result that is completely consistent with
independent analysis using the 2dFGRS \citeauthor{2pigg}.  We also see
that 25$\%$ of L$_*$ galaxies are in relatively large groups or
clusters with halo masses above 10$^{13}$ \Mdoth. This is much larger
than the $\sim$ 10 $\%$ claimed by B09, likely a result of the way
they assign galaxies to subhalos, as discussed further in \textsection
\ref{sec-previous}. In particular, B09 assign a galaxy to a subhalo if
the subhalo mass is $> 10^{11.5}$ \Mdoth\ when it is accreted into a
more massive host. However, the mass in a subhalo begins to be tidally
stripped significantly before reaching the virial radius of a more
massive host, even without significantly disturbing the galaxy within
\citep{natarajan}.

\begin{table*}
\begin{tabular}{| c | c | c | c | c | }
\hline
Redshift & Number of clusters & Mass range  & Median mass  & Average
number of \\
 & & Log(\Mdoth) &  Log(\Mdoth) & galaxies per cluster \\
\hline
 0  & 40 & 15.0-15.6 & 15.14 & 1161  \\
    & 189 & 14.7-15.0 & 14.82 & 569 \\
    & 673 & 14.4-14.7 & 14.53 & 297 \\
    & 1822 & 14.1-14.4 & 14.24 & 156 \\ 
    & 4404 & 13.8-14.1 & 13.94 & 78 \\
    & 9325 & 13.5-13.8 & 13.64 & 41 \\
    & 18730 & 13.2-13.5 & 13.34 & 20 \\
    & 36265 & 12.9-13.2 & 13.04 & 10 \\

\hline
0.5 & 4 & 15.0-15.6 & 15.16 & 1161  \\
    & 29 & 14.7-15.0 & 14.79 & 536  \\
    & 212 & 14.4-14.7 & 14.51 & 289 \\
    & 786 & 14.1-14.4 & 14.23 & 156 \\ 
    & 2471 & 13.8-14.1 & 13.93 & 80 \\
    & 6325 & 13.5-13.8 & 13.68 & 42 \\
    & 14440 & 13.2-13.5 & 13.34 & 22 \\
    & 30124 & 12.9-13.2 & 13.04 & 11 \\
\hline
1.0 & 0 & 15.0-15.6 & -- & --  \\
    & 3 & 14.7-15.0 & 14.82 & 532  \\
    & 40 & 14.4-14.7 & 14.51 & 252  \\
    & 275 & 14.1-14.4 & 14.22 & 137 \\ 
    & 1134 & 13.8-14.1 & 13.92 & 72 \\
    & 3643 & 13.5-13.8 & 13.63 & 38 \\
    & 9820 & 13.2-13.5 & 13.34 & 21 \\
    & 23388 & 12.9-13.2 & 13.04 & 11 \\
\hline
1.5 & 0 & 15.0-15.6 & -- & --  \\
    & 1 & 14.7-15.0 & 14.81 & 381  \\
    & 2 & 14.4-14.7 & 14.41 & 178  \\
    & 55 & 14.1-14.4 & 14.19 & 119 \\ 
    & 322 & 13.8-14.1 & 13.92 & 66 \\
    & 1528 & 13.5-13.8 & 13.62 & 35 \\
    & 5465 & 13.2-13.5 & 13.33 & 19 \\
    & 15134 & 12.9-13.2 & 13.03 & 10 \\
\hline

\end{tabular}
\caption{Properties of the cluster sample derived from
  \citet{font}. The first column lists the redshift snapshot from
  which the clusters were selected and the second column gives the
  total number of clusters used for analysis in each bin. Columns 3
  and 4 list the cluster halo mass range and median mass of clusters
  in that range. We use these halo mass bins extensively in the rest
  of the paper. Column 5 lists the average number of galaxies per
  cluster with stellar masses above $M$ = 10$^{9}$ \Mdoth\ at the epoch
  of observation.}
\label{table-clusterprops}
\end{table*}

\section{Results} \label{sec-results}

We now look in detail at how the cluster galaxies end up in the
clusters, and what insights this might give into the processes which
might affect those galaxies.

\subsection{Cluster and group accretion history}

Galaxies which have been in massive halos prior to joining the final
environment may have been environmentally pre-processed. Thus, we
begin by examining the host halo masses of galaxies just prior to
their accretion into the final group or cluster halo.  To acheive
this, we trace the most massive progenitor of every galaxy, back
through each snapshot in the simulation. We record the halo mass of
this progenitor in the timestep just before it becomes a member of the
final cluster, which defines its environment at the time of accretion.

\begin{figure}
\leavevmode \epsfysize=8cm \epsfbox{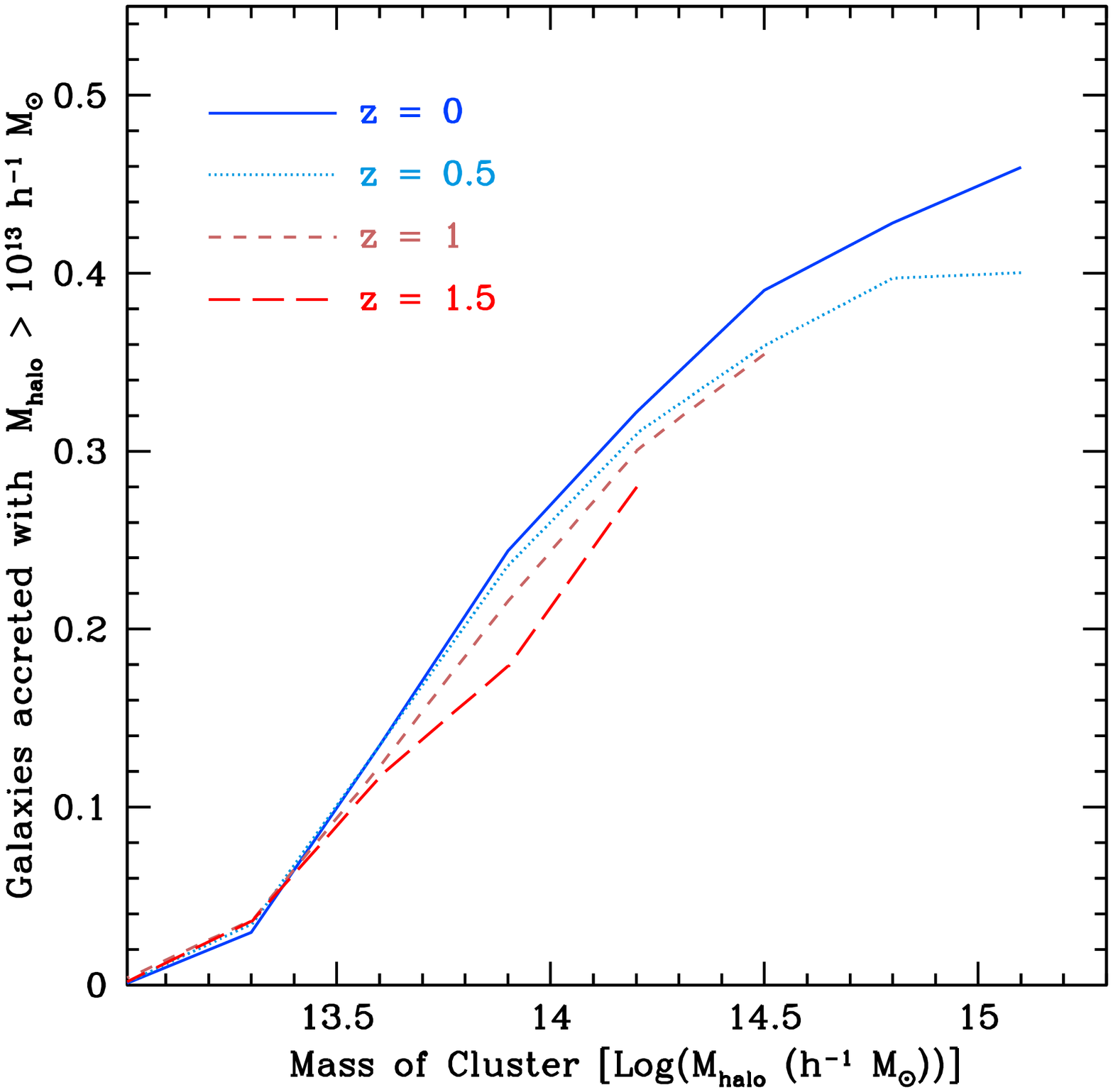}
\caption{The fraction of cluster galaxies which were accreted into the
  final cluster halo as a member of a halo with $M>10^{13}$ \Mdoth.  This is
  shown as a function of the final cluster mass at the epoch of
  observation, for four redshifts.  All cluster galaxies have final
  stellar masses of M $>$ 10$^{9} M_\odot$. The mass range bins were
  defined in Table 1, and are shown for all bins containing more than
  two clusters.}
\label{gal_accr_sim}
\end{figure}

We show the full accretion histories for all the cluster mass bins, in
each of the four redshift epochs, in the Appendix. Here we will examine
the most important insights which can be drawn from those accretion histories.
Figure \ref{gal_accr_sim} shows the fraction of galaxies in the final
cluster which were accreted through haloes at least as massive as
10$^{13}$ \Mdoth\ (large groups). We show this as a function of the final
cluster mass for each of four redshift epochs. We first consider
relatively low-mass clusters, with $M\sim 10^{14.2} M_\odot$ at z=0.
We find that 32 per cent of galaxies in these clusters were accreted through such group-sized
halos.  This is somewhat higher than the 24 per cent found by B09; the
small difference can be related to the difference in the way haloes are
populated with galaxies, as we discuss in \S~\ref{sec-previous}.

However, such clusters are fairly poor systems; they are less massive
than all 16 clusters observed extensively by the CNOC1 collaboration
\citep{cnoc_masses}, and an order of magnitude smaller than the nearby
Coma cluster (M$_{200}$ = 1.88 $^{+0.65} _{-0.56}$ $\times$ 10$^{15}$
\Mdoth, \citep{kubo}).  Figure \ref{gal_accr_sim} shows that the
fraction of galaxies which are accreted through group sized halos is
strongly dependent on the mass of the final halo. This is because
massive haloes are not surrounded by an average patch of the universe,
but tend to be strongly clustered with other massive halos
\citep[eg.][]{Kaiser}. At z=0, we see that 45 $\%$ of galaxies
accreted into a cluster of Coma's mass have been accreted from haloes
with $M>10^{13}$ \Mdoth.  This suggests that pre-processing in group
environments before cluster accretion may be significant.
Interestingly, the fraction of galaxies accreted through massive
haloes has only a weak dependence on the redshift of observation of
the cluster. In other words, a Coma-sized cluster at z=0.5 would
accrete 40 $\%$ of its galaxies from $M>10^{13}$ \Mdoth\ halos. The
galaxy assembly histories are remarkably similar, with the dominant
difference being simply that Coma-sized clusters do not exist in the
relatively small volume of the Millennium simulation at $z= 1.0$ or
$1.5$.

\begin{figure}
\leavevmode \epsfysize=8cm \epsfbox{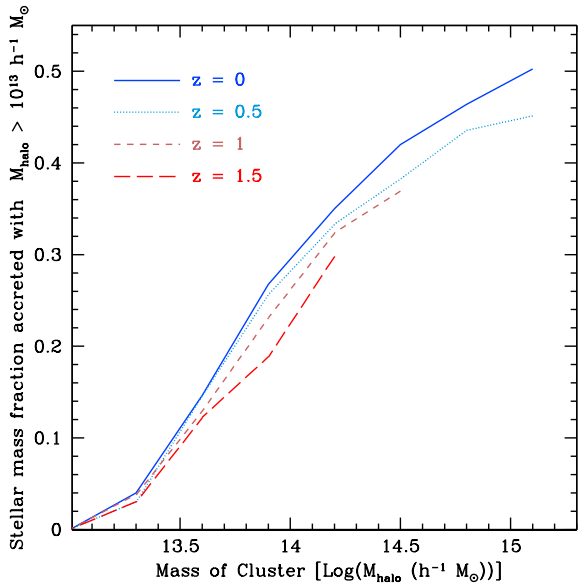}
\caption{As Figure~\ref{gal_accr_sim}, but showing the fraction of
  accreted {\it stellar mass} which resides in a $M>10^{13}$ \Mdoth\
  halo at the time of accretion.  }
\label{stel_accr_sim}
\end{figure}

\begin{figure}
\leavevmode \epsfysize=8cm \epsfbox{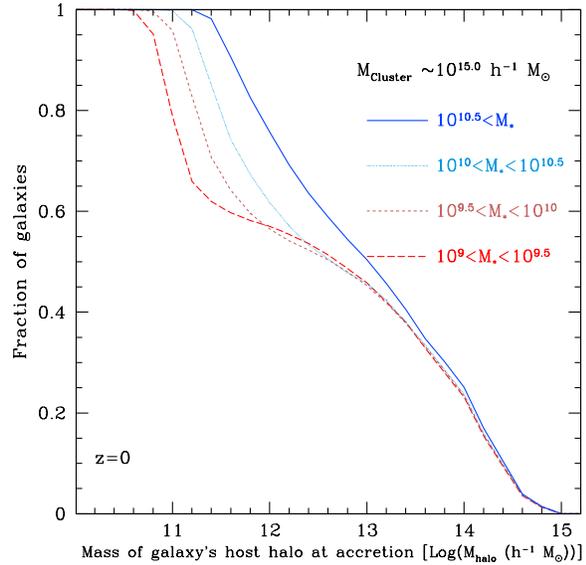}
\caption{
The cumulative distribution of accreted cluster galaxies as a
  function of host halo mass at the time of accretion into the final
  cluster. The distribution is shown in three stellar mass bins at
  z=0, for a final cluster with $M=10^{15}$\Mdoth.}
\label{lumdep_accr}
\end{figure}

In Figure \ref{stel_accr_sim}, we show the fraction of stellar mass
which is accreted through halos at least as massive as 10$^{13}$
\Mdoth. This figure is quite similar to Figure
\ref{gal_accr_sim}. However, notice that the fraction of {\it stellar
mass} accreted by the most massive clusters through groups is larger
than the fraction of {\it galaxies} accreted through such systems.
Indeed, the stellar mass accretion history closely matches the
expected behavior of the dark matter accretion. The extended Press
Schechter formalism and n-body simulations of dark matter roughly
agree that $\sim$ 30 $\%$ of the mass of a halo is accreted from halos
with masses a tenth the mass of the final halo \citep{bond_eps,
bower_eps, lacey_cole, stewart}. We find this same fraction for all
our stellar mass accretion histories, while the fraction of galaxies
accreted is smaller at high cluster mass.  This implies there are are
fundamental differences in how galaxies are accreted as a function of
their stellar mass.  This is illustrated in Figure \ref{lumdep_accr},
where we show the accretion histories of galaxies which end up in
a$M=10^{15.0}$ \Mdoth\ cluster at z=0, binned by their final stellar
mass.  There is a large difference in the masses of the host halos
prior to accretion for low and high mass galaxies. While $\sim$ 52
$\%$ of the most massive galaxies are accumulated from haloes with
$M>10^{13}$ \Mdoth, this is only the case for $\sim$ 45 $\%$ of the
least massive galaxies we consider.  This discrepancy is much larger
if we consider accretion through poorer groups, with $M>10^{12}$
\Mdoth.  The more massive galaxies are more likely to have been
accumulated from group mass halos, and thus more likely to have been
pre-processed prior to accretion into a cluster.

\begin{figure}
\leavevmode \epsfysize=8cm \epsfbox{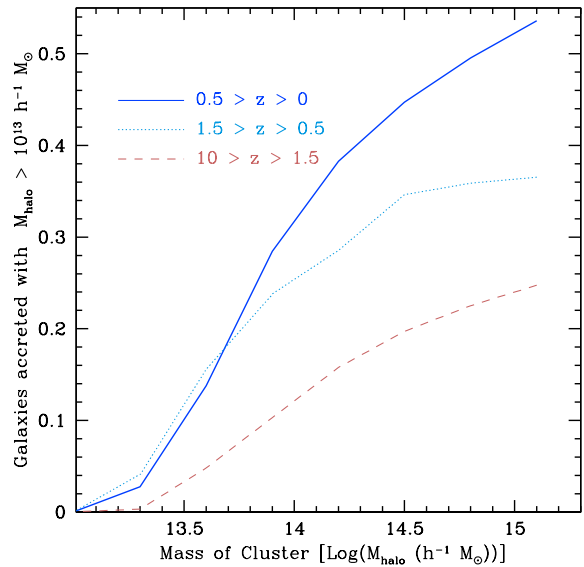}
\caption{The fraction of cluster galaxies which were accreted into the
  final cluster halo as a member of a 10$^{13}$ \Mdoth\ halo or
  greater.  This is shown as a function of the final cluster mass z=0
  and for three bins in accretion redshift.  All cluster galaxies have
  final stellar masses of M $>$ 10$^{9} M_\odot$. }
\label{gal_accr_time}
\end{figure}

Now that we have seen that the degree of group preprocessing depends
on both the stellar mass of the galaxy and the mass of the final
cluster, we would like to examine how this varies as a function of
redshift. In Figure \ref{gal_accr_time} we show the fraction of
cluster galaxies which were accreted into the final cluster halo as a
member of a halo with $M>10^{13}$ \Mdoth\ halo. This is broken up into
three bins, which represent the redshift at the time of the galaxy's
accretion into the cluster. From this we see that the degree of
preprocessing is significantly dependent on the time the galaxies were
accreted. Galaxies which are accreted recently into the cluster are
more likely to have been in a group environment than ones accreted
into the cluster at high redshift. In particular, since $z=0.5$ the
most massive clusters today have accreted most of their new galaxies
via infalling groups.

\subsection{Cluster and group assembly histories}

We have seen that the accretion history of clusters varies with final
cluster mass, is a function of the stellar mass of the accreted
galaxy and is dependent on the redshift of accretion. However, this
does not address the state of the cluster itself. The importance of
pre-processing depends not only on the accretion history but also on
the amount of time the main cluster progenitor itself had the mass of
a group.

\begin{figure*}
\includegraphics[width=\textwidth]{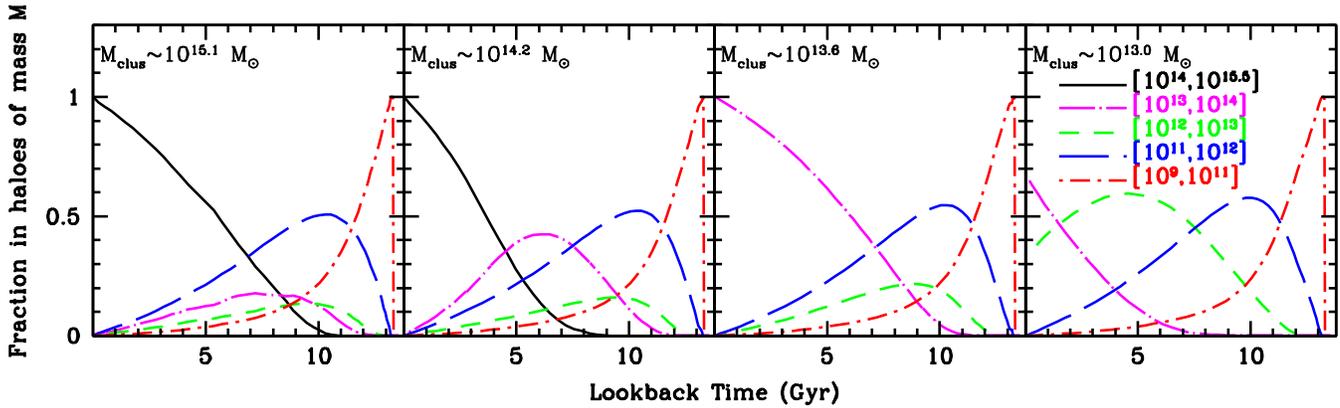}
\caption{The fraction of galaxies residing in z=0 clusters that are
  found in halos of mass $M$ at a previous time $t$. We only consider
  the most massive progenitor of each cluster galaxy. Each curve shows
  a range of M, indicated by the legend, in units of \Mdoth. Each
  panel represents clusters of different final masses, as indicated in
  the top-left corner. Note that the curve corresponding to haloes
  with $10^{11}<M/$\Mdoth$<10^{12}$ is very similar in all cases,
  indicating that the accretion rate of ``isolated'' galaxies is
  roughly independent of final cluster mass.  }
\label{gal_loc}
\end{figure*}

Therefore, to get a complete picture of the assembly of galaxy
clusters and groups and the halo masses which are important for the
properties of their galaxies we present Figure \ref{gal_loc}. This
shows the distribution of halo masses in which the most massive
progenitors of final z=0 cluster galaxies reside, as a function of
lookback time and for four bins of final z=0 cluster mass.  The panels
in Figure \ref{gal_loc} show distinctly different assembly histories
for very massive clusters, smallish clusters, and groups. In
particular, the relative importance of the group environment varies
tremendously for these three types of structures. The most massive
cluster never has more than 17 $\%$ of galaxies in group sized halos
(10$^{13}$ \Mdoth $<$ $M_{\rm halo}$ $<$ 10$^{14}$ \Mdoth) while as
many as 44$\%$ of the galaxies in a $M\sim10^{14.2}$ \Mdoth\ cluster
today have spent some time within such haloes in the past.  In fact,
for a period of 2 Gyrs, groups are the most common environment of the
galaxy progenitors; this is because during this time the main cluster
progenitor itself has the mass of a group.  Thus, considering only the
haloes of galaxies prior to accretion into the main cluster may
underestimate the role of the group environment, as already noted by
B09.

Given the distinctly different assembly histories of these clusters
and massive groups, it is perhaps surprising that observations of
large samples of galaxy clusters in the local universe show that the
fraction of red galaxies is approximately constant in clusters more
massive than 10$^{13.8}$ \Mdoth\ \citep{Hansen}. Therefore, it is
useful to look for some common trait in the assembly histories of
clusters which may point to the reason for this uniformity. It is
interesting that the population of 'isolated' galaxies, those in
10$^{11}$ \Mdoth $<$ $M_{\rm halo}$ $<$ 10$^{12}$ \Mdoth, shows a
similar distribution in the four different panels. At a lookback time
of 10 Gyrs, $\sim$ 55 $\%$ of cluster galaxy progenitors were in this
halo mass regime, and that percentage has declined at a nearly
constant rate of 5 -- 6 $\%$ per Gyr until the current epoch,
regardless of the final cluster mass. In other words, the distribution
of galaxies not in 'isolated' halos is similar regardless of final
cluster mass, and supports the hypothesis that the galaxy
transformation mechanism begins to occur as galaxies leave their
'isolated' halos.

\begin{figure*}

\includegraphics[width=\textwidth]{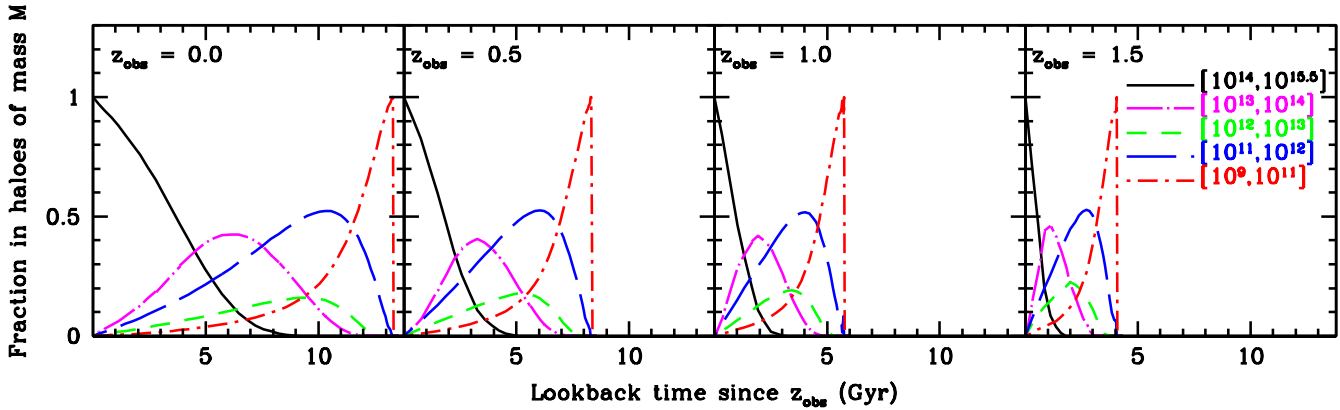}
\caption{As in Figure \ref{gal_loc}, but for a cluster of mass
  10$^{14.5}$ (\Mdoth) observed at four redshift epochs as indicated
  at the top of each panel.  Clusters of a given mass at higher
  redshift must assemble their mass more quickly, and thus the time
  available for pre-processing through group-sized haloes is
  decreased.  }
\label{gal_loc_z}
\end{figure*}

Finally, we examine the assembly histories of galaxy clusters
of a given mass at each redshift epoch. In Figure \ref{gal_loc_z} we
show the distributions of halo masses for the most massive progenitors
of 10$^{14.5}$ \Mdoth\ cluster galaxies as a function of lookback time
at all four redshift epochs. While the final cluster mass is the same
(at each epoch), the higher redshift clusters must assemble their mass
more quickly and thus their galaxies have not been in massive halos
for as long. For instance, 5 Gyrs prior to the observation epoch,
$\sim$ 50$\%$ of z=0 galaxies were in 10$^{14}$ M$_\odot~h^{-1}$
haloes, while none of the z=1 or z=1.5 cluster galaxies were even in
10$^{13}$ \Mdoth\ haloes yet.  Environmental processes have had a much
longer timescale over which to affect low redshift groups and clusters
than higher redshift ones.

There are two interesting points when comparing Figure \ref{gal_loc_z}
with Figure \ref{gal_loc}. First, we see that the maximum fraction of
galaxies in each halo mass bin is the same in clusters of the same
final mass but seen at different redshifts. For instance, the maximum
fraction of galaxies which reside in halos of 10$^{13}$ \Mdoth $<$
$M_{\rm halo}$ $<$ 10$^{14}$ \Mdoth\ at any time is 40 $\%$ regardless of
the redshift epoch. The lookback time at which these maximum fractions
occur varies significantly with redshift, but it would appear their
path through the hierarchy is similar. Essentially, clusters of fixed
mass at different redshift epochs have assembly histories which become
more stretched out at lower redshift. The assembly histories would
look almost identical if the lookback time was divided by the age of
the universe at that redshift epoch. This result was hinted at in
Figure \ref{gal_accr_sim}, which showed that the fraction of galaxies
accreted through massive haloes was approximately the same at all
redshift epochs for a cluster of given mass.

This leads to the second interesting observation to be made from
Figure \ref{gal_loc_z}. The rate at which galaxies leave their
'isolated' halos increases significantly with redshift. At z=0, as
before, for the 10 Gyrs prior to observation the fraction of galaxies
in halos of 10$^{11}$ \Mdoth $<$ $M_{\rm halo}$ $<$ 10$^{12}$ \Mdoth
decreases by about 5-6$\%$ per Gyr, while 10$\%$ (15$\%$) [20$\%$] of
galaxies leave their 'isolated' halos per Gyr at a constant rate for
5(3.5)[2.5] Gyrs prior to observation at z=0.5(1)[1.5]. Therefore, the
accretion rate of galaxies from isolated environments into groups and
clusters is higher at higher redshift. Again, this result is a direct
result of the reduced time between the epoch of observation and the
beginning of the universe. The assembly histories at higher redshift
are just compressed, leading to a higher accretion rate, even
though the total number accreted from isolated environments is
constant at each epoch of observation. The effect this has on the
galaxy properties of galaxy clusters as a function of redshift will be
discussed in the following section.

\subsection{Cluster to cluster variation in environmental effects} \label{simred}

We have established the galaxy accretion history and galaxy assembly
history of galaxy clusters at a range of epochs. We would now like to
assess how these galaxy histories affect the final properties of
galaxies at each redshift epoch. To this end, we examine the fraction
of galaxies in each cluster, which have been within dense environments
long enough to expect that environmental effects might be
important. By examining the fraction of environmentally affected
galaxies in each cluster, we can quantify both the total numbers of
affected galaxies, and their variation from cluster to cluster.

In a simple way, we can parametrize the length of time it takes for a
galaxy to display an environmental effect, \Ttrunc, after falling into
a halo with a mass above a characteristic mass threshold,
\Mtrunc. Although it is not obvious that there is a single main
physical mechanism which causes the environmental affects displayed in
both groups and clusters, we explore the predictions of such a model
and discuss the limitations of this approach in the following section.

\begin{figure*}
\leavevmode \epsfysize=16cm \epsfbox{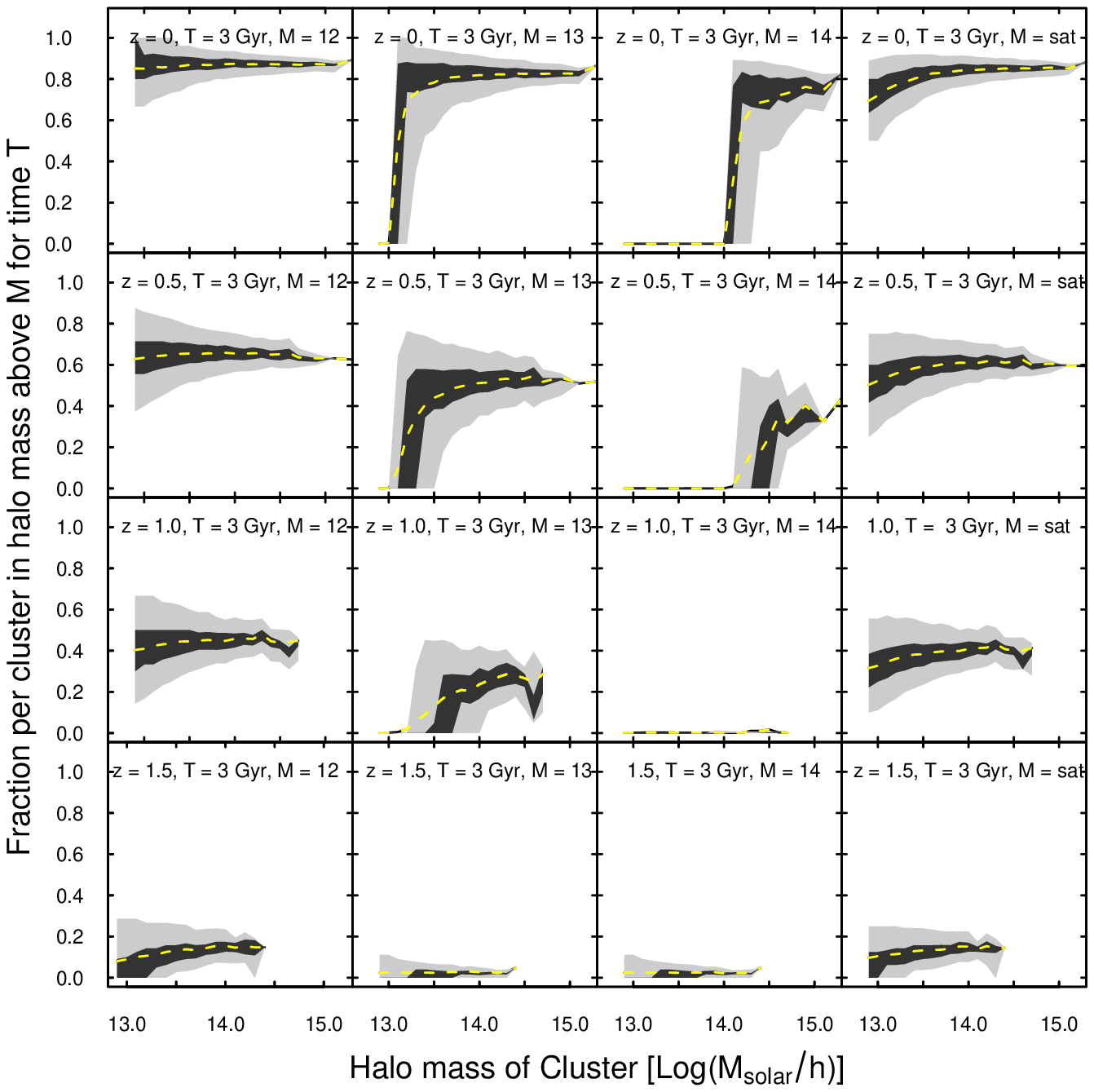}
\caption{The fraction of cluster galaxies with $M>10^9$\Mdoth\ that have
  resided within a halo of mass $M\geq$\Mtrunc\ for a time
  $t\geq$\Ttrunc\ is shown as a function of final cluster mass.  We
  interpret this as the fraction of ``environmentally--affected''
  population in our simple model.  The panels contain four contour
  lines marking the 10, 33, 67 and 90 percentiles of the distribution
  in this fraction, while the dashed yellow line represents the
  average.  The truncation time is fixed at \Ttrunc$=3$ Gyr, and each
  row shows a different assumption for \Mtrunc, as indicated.
  Different rows correspond to clusters at a different redshift, as
  indicated. }
\label{predict_scatter}
\end{figure*}

\begin{figure*}
\leavevmode \epsfysize=16cm \epsfbox{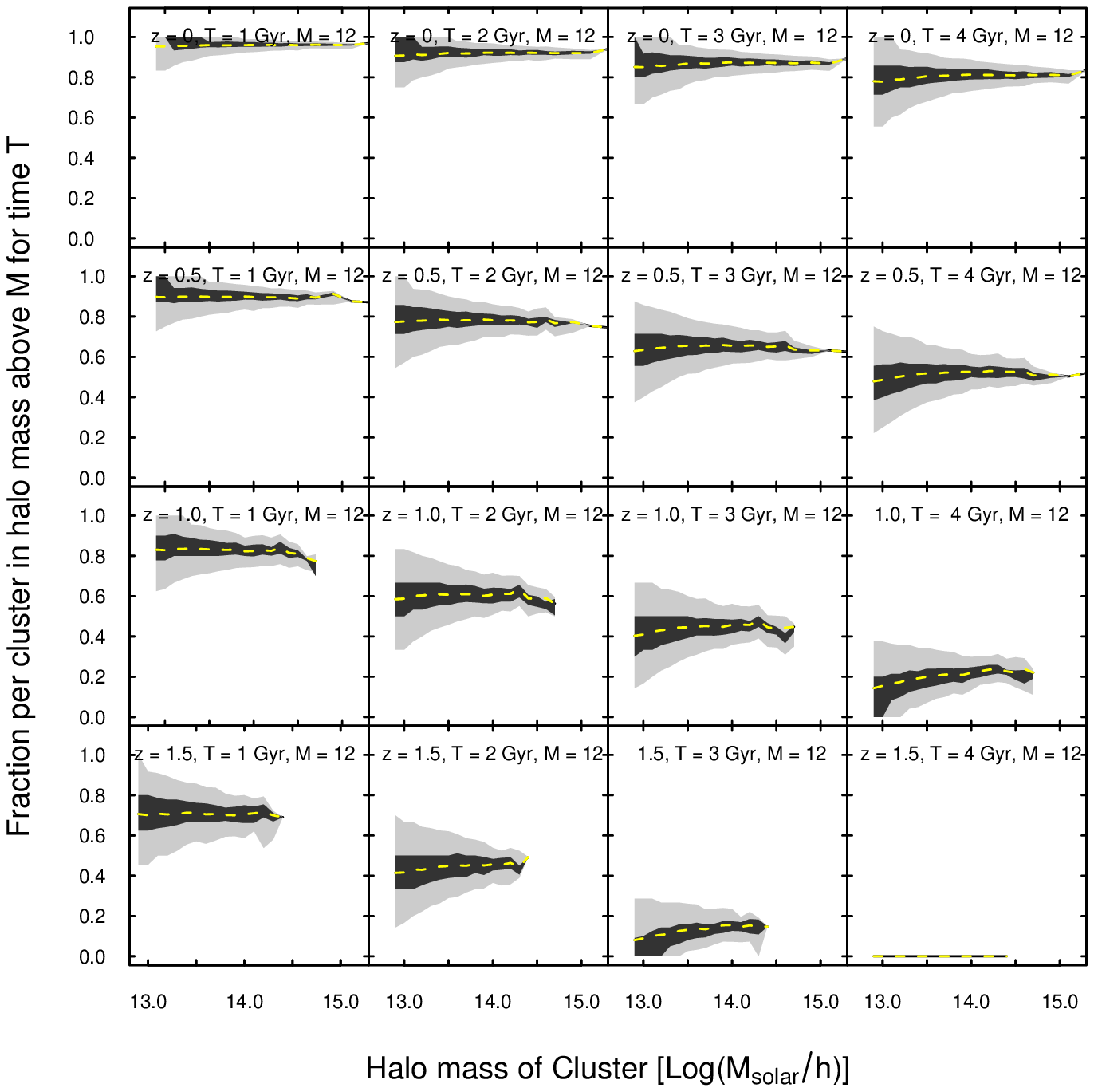}
\caption{ As Figure~\ref{predict_scatter}, but where the
characteristic halo mass threshold is fixed at \Mtrunc$=10^{12}$
\Mdoth, and the truncation times \Ttrunc\ are varied along rows of the
figure, from 1--4 Gyr as indicated.  }
\label{predict_scatter_time}
\end{figure*}

Given this model we can explore how varying the truncation time,
\Ttrunc, and the characteristic mass threshold, \Mtrunc, alters the
implied environmental effects on galaxies. In Figure
\ref{predict_scatter}, we show the predicted average fraction of
galaxies in each cluster which are subject to environmental effects in
our simple model; the distribution of this fraction is reflected in
the four contour lines marking the 10, 33, 67 and 90 percentiles.
Here we fix the truncation time, \Ttrunc, to be 3 Gyr and allow the
characteristic mass threshold, \Mtrunc, to vary from 10$^{12}$ \Mdoth\
to 10$^{14} M_\odot$. In other words, in this figure, a galaxy has
felt an ``environmental effect'' if it has been within a halo of mass
$M\geq$\Mtrunc\ for at least 3 Gyrs. In addition, we allow a fourth
category, in which the expression of an environmental effect occurs 3
Gyrs after the galaxy has become a satellite galaxy in a larger dark
matter halo, regardless of its mass.

This figure has some noteworthy features. First, for massive clusters
(M $>$ $10^{14.5}$ \Mdoth) at z=0, the mean number of environmentally
affected galaxies in this model is similar ($\sim$ 80-85 $\%$)
regardless of \Mtrunc. The implication of this for low redshift
observational studies is that it is difficult to discern the value of
the characteristic mass threshold by observing systems above that mass
threshold. This highlights the importance of studies of low mass
galaxy groups. Observations at low and intermediate redshift show that
group galaxies with a given stellar mass have properties distinct from
the average field galaxy; if our simple model of environment-driven
transformation is correct, this indicates a characteristic mass
threshold of at least this scale (M $\approx$ 10$^{12.5}$ - 10$^{13}$
\Mdoth) \citep{wilmancnoc, weinmann}.

Indeed, as previously mentioned, low redshift observations show that
the fraction of red galaxies in clusters is essentially uniform, for
clusters with $M>10^{13.8}$ \Mdoth\ \citep{Hansen}.  Given this, it is
also worth noting that in Figure \ref{predict_scatter}, our model also
produces a strikingly flat fraction of environmentally affected
galaxies per cluster as a function of cluster mass. This is a direct
result of the behavior noted in Figure \ref{gal_loc}, that the
fraction of galaxies infalling from isolated halos is independent of
halo mass.

Although it may be difficult to use the average properties of massive
clusters at a given epoch to discern the characteristic mass
threshold, one possible method would be to observe the variation in
their properties. The predicted scatter in the fraction of
environmentally affected galaxies per cluster is quite small ($\sim$ 5
$\%$) for 10$^{14.5}$ \Mdoth\ clusters at z=0 when \Mtrunc = 10$^{12}$
\Mdoth, but close to 40 $\%$ when \Mtrunc = 10$^{14}$ \Mdoth. The
scatter in, for instance, the fraction of early type galaxies or
optical line emitting galaxies in clusters at z=0 is much smaller than
40 \% \citep{dressler80,poggianti,finn}. We will examine the scatter
in red fractions of galaxies in clusters at z=0 in a future
paper. Unfortunately, the scatter at z=0 of a model where \Mtrunc =
10$^{12}$ \Mdoth\ is not that different from a model where \Mtrunc =
10$^{13}$ \Mdoth. However, notice that the scatter in these two models
becomes more significant at z $>$ 0. Intriguingly, \citet{dressler}
showed that, while the morphology-density relation was equally strong
in all clusters at low redshift, the relation was stronger in
centrally--concentrated clusters than irregular clusters at z $\sim$
0.5.\footnote{While the Dressler et al. results, and many intermediate
redshift results, have limiting stellar mass on the order of 10$^{10}$
\Mdoth compared with our limit of 10$^{9}$ \Mdoth, we have verified
that the scatter in the cluster red fractions is constant with a
limiting mass change to 10$^{10}$ \Mdoth.} In effect, this suggests
that the scatter in the fraction of environmentally affected galaxies
of each cluster is significant at z $\sim$ 0.5. Although not
definitive, this may point to a characteristic mass threshold which is
somewhat larger than 10$^{12}$ \Mdoth, given that scatter in that
model is still quite small at z=0.5 ($\sim$ 13 $\%$ at 10$^{14.5}$
\Mdoth). Notice that a model where the environmental effects begin to
occur when a galaxy becomes a satellite behaves very similarly to a
model with \Mtrunc = 10$^{12}$ \Mdoth. We discuss this similarity
further in \textsection \ref{sec_sem}.

Examining the redshift evolution of any of the given models shows that
they all predict a significant Butcher-Oemler effect. That is, they
predict that there are fewer environmentally affected galaxies in
clusters with increasing redshift. In particular, by z= 1.5 all of the
models predict a very small or non-existent fraction of
environmentally affected galaxies. Indeed, the 10$^{14}$ \Mdoth\ model
leads to the prediction that, by $z = 1$, no galaxies will be
environmentally affected.

Our choice of \Ttrunc = 3 Gyrs in the models presented above is ad
hoc, and we would like to quantify how changing the timescale effects
the predictions. In Figure \ref{predict_scatter_time}, we explore a
model in which the characteristic halo mass, \Mtrunc, is kept fixed at
10$^{12}$ \Mdoth, and allow \Ttrunc to vary from 1 Gyr to 4 Gyrs. We
show the fraction of environmentally affected galaxies for each of the
four redshift epochs of our clusters. Although \Mtrunc\ is held
constant, we note that the results and our interpretation are similar
for any choice of \Mtrunc\ within the range $10^{12}$--$10^{13}$
\Mdoth, which seems the most likely value given the arguments above.

Similarly to Figure \ref{predict_scatter}, for each \Ttrunc, we see a
significant Butcher-Oemler effect, such that clusters at higher
redshift have fewer galaxies affected by environmental
processes. However, the size of the effect even between $z=0$ and
$z=0.5$ is dramatically altered by the choice of time scale.  With a
short timescale of only 1 Gyr, the fraction of
environmentally-affected galaxies evolves little, from $\sim$85\% at
$z=0.5$ to $\sim 95$\% today.  On the other hand, a long timescale of
\Ttrunc$=4$ Gyr results in a much stronger evolution over this
redshift range, from 50\% to 80\%.  Compare this evolution with that
observed in the red fraction of cluster galaxies, which indicate an
evolution of $\sim$ 25 $\%$ over a similar redshift range, from 0.9 at
z=0.2 to 0.65 at z=0.5 \citep{ellingson}. This seems to indicate that
a relatively long time scale for the expression of environmental
effects ($>$ 2 Gyr) would be required to match this quick evolution. A
similar timescale is necessary to explain the radial gradient of
passive galaxies in galaxy clusters \citep{ellingson, balogh_model}.

The predicted scatter from cluster to cluster is also
noteworthy. Recall that in Figure~\ref{predict_scatter} we saw that
the scatter was sensitive to the characteristic halo mass used. In
this plot, for the majority of the time, the scatter is similar at
each redshift regardless of the timescale for truncation. This
strengthens our previous argument that a well-defined measure of the
scatter in cluster properties at a given redshift could allow one to
discern the characteristic halo mass for truncation.

We have provided strong evidence, which we summarize in \textsection
\ref{sec_obser}, that the dominant environmental processes at work in
galaxy groups and clusters begin to become effective at a halo mass
scale of 10$^{12}$ - 10$^{13}$ \Mdoth, and are active for a timescale
of at least a few Gyrs. Given these constraints, we see that figure
\ref{predict_scatter_time} predicts that by z=1.5 there should be
little to no environmental effect on galaxies. Remarkably, this
prediction has some observational evidence to suggest it is
correct. \citet{cooper} showed, using galaxies selected from the DEEP2
redshift survey, that the red fraction only weakly correlates with
overdensity at z $\sim$ 1.3. While the comparison to our predictions
is complicated because the Cooper et al. sample only includes massive
galaxies, this is not a trivial agreement; in fact, assuming the
timescale was 1 Gyr, this would lead us to predict that 70 $\%$ of
galaxies at z=1.5 are still environmentally affected. This fraction
would be even higher when we used the same limiting stellar mass as
Cooper et al., as is discussed in the next section. Additionally, the
DEEP2 survey is complicated by their rest frame blue magnitude limit
which causes them to naturally detect fewer and fewer red galaxies at
higher redshifts. Further, DEEP2 does not cover a wide enough area to
have massive clusters within it, so targeted, stellar mass limited
studies of the extreme cluster environments are still needed at this
redshift to quantify the size of the environmental effects. Although,
given our results, they may be difficult to find using the popular and
efficient red sequence method \citep{gladders, Lu}.

\subsection{Stellar mass dependence of environmental effects} \label{sec_galprop}

\begin{figure}
\leavevmode \epsfysize=8cm \epsfbox{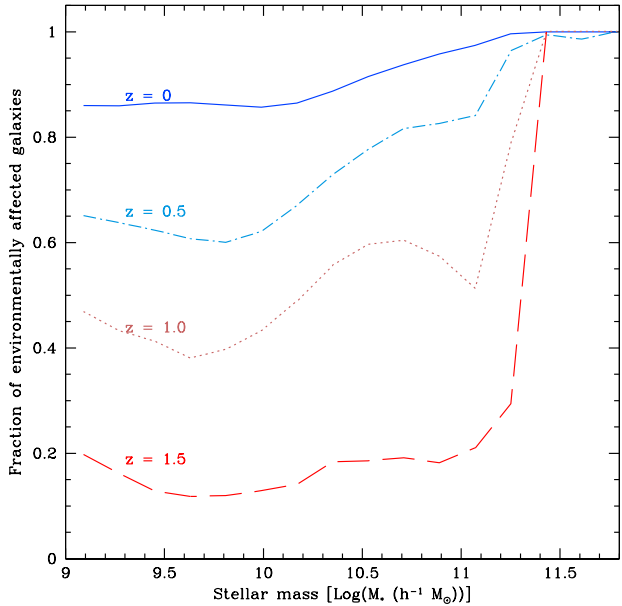}
\caption{The fraction of environmentally affected galaxies in clusters
  at all four redshift epochs as a function of galaxy stellar
  mass. Cluster galaxies have host halo masses greater than
  10$^{14}$ \Mdoth at the epoch of observation. This model assumes M$_{trunc}$=10$^{12}$\Mdoth\ and \Ttrunc =
  3 Gyrs as in Figure \ref{predict_scatter}. }
\label{gal_redstel}
\end{figure}

Observations suggest that the fraction of galaxies which are passive
or red, depends greatly on their own stellar mass \citep{baldry06,
haines}. It is thought that this is at least partially due to secular
influences, ie. AGN feedback, which primarily occur in massive
galaxies \citep{kauffmann_agn}. We can use our simple model for
environmental effects to examine the fraction of cluster galaxies
(those with $M_{halo} $ $>$ 10$^{14}$ \Mdoth)  
which may also be subject to environmental effects. This is presented
in Figure \ref{gal_redstel} for a model which has \Mtrunc = 10$^{12}$
\Mdoth\ and \Ttrunc $= 3$ Gyr at all four redshift epochs.

The fraction of environmentally affected galaxies is a strong function
of stellar mass in this model, and the gradient becomes stronger with
increasing redshift.  The most massive galaxies have resided within
group-sized haloes since at least $z=1.5$; thus any environmental
effects would have manifested themselves a long time prior to
observation, and we expect to see little signature of cluster growth
in their properties.  On the other hand, galaxies with lower stellar
mass are a better tracer of the recent mass accretion history of the
cluster, hence we see a strong evolution in the fraction of
environmentally--affected galaxies.

\subsection{Observational constraints} \label{sec_obser}

It is now useful to review the observational constraints on our model
parameters, \Mtrunc and \Ttrunc. The halo mass threshold at which
environmental effects become important must be at least as low as
10$^{13}$ \Mdoth\ because there are observations of systems at this
mass with significant environmental effects \citep{wilmancnoc,
weinmann}. For this reason, we investigated a model with a low halo
mass threshold, 10$^{12}$ \Mdoth, in Figure
\ref{predict_scatter_time}. Well defined samples of galaxy clusters
show a significant Butcher-Oemler effect, such that the fraction of
red galaxies decreases from $\sim$ 0.9 at z=0.2 to $\sim$ 0.65 at
z=0.5 \citep{ellingson}. This evolution is much quicker than predicted
by a model with a \Ttrunc of 2 Gyrs or less. Thus a model with \Mtrunc
of $\sim$ 10$^{12}$ \Mdoth\ and a \Ttrunc of $\sim$ 3 Gyrs is the most
favored model. As suggested previously, this leads to the prediction
that by z=1.5, little or no environmental effects are felt by the
galaxy population.

Recall \textsection \ref{sec_galprop}, in which we investigated the
stellar mass dependence of the galaxy population using our most
favored model. We found that while the most massive galaxies are
environmentally affected at all redshifts, the lower mass galaxies
become more affected with time. \citet{gilbank} used a compilation of
the observational literature to show that the ratio of red bright
galaxies to red faint galaxies steadily increases with redshift, the
same qualitative behaviour we see in the simple model.

It is difficult to observationally quantify the extent to which
massive galaxies are environmentally affected. This is largely because
the visual colours of galaxies are not very sensitive to low levels of
star formation. Mid-IR observations are more sensitive to low levels
of star formation and thus are better at establishing the
environmental influence of massive galaxies. Observations at z $\sim$
0.4 suggest that only 10 $\%$ of massive galaxies ($>$ 10$^{10}$
\Mdoth) in groups have IR emission indicative of activity, while the
global fraction is much higher \citep[$\sim
40$\%,][]{wilmanIRAC}. Additionally, \citet{wolf} find that massive
galaxies are uniformly old and red in the cluster cores, while having
a significant population of dusty, star-forming red galaxies in the
infall regions. Both of these studies suggests that significant
environmental effects are felt even by massive galaxies, as assumed in
our model.

We emphasize that this observational comparison is qualitative, yet
highly suggestive. In a future paper we investigate the quantitative
behavior of these models with a direct comparison to the best
available cluster, group and field data to z$\sim$ 1.

\subsection{Comparison to Previous Work} \label{sec-previous}

In an attempt to explain observations of the fraction of cluster
members with [OII] emission at z=0 and z=0.6, \citet{poggianti} have
presented a similar, but more complex model. The observations they
present (their Figure 4) show that while, at z=0.6, higher mass
clusters have lower average fractions of [OII] emitting galaxies, this
is largely because of an upper envelope which decreases with
increasing cluster velocity dispersion. In contrast, they notice that
at z=0, the fraction of [OII] emitting galaxies is constant with
cluster velocity dispersion above 550 km/s ($\sim$ 10$^{14}$ \Mdoth),
but the scatter is large below that value.

In effect, to explain the observations, \citet{poggianti} introduces
two \Mtrunc\ and two \Ttrunc\ parameters to match the observed
behavior. The first set of \Mtrunc\ and \Ttrunc\ are meant to
represent 'primordially' passive galaxies, and are associated with
elliptical galaxies. They claim that galaxies within
3$\times$10$^{12}$ \Mdoth\ groups at $z=2.5$ represent these
primordially passive galaxies. The second set of parameters are
associated with quenched galaxies or S0 galaxies, and are set to have
\Mtrunc=10$^{14}$ \Mdoth\ and \Ttrunc$=3$ Gyrs. However, observations
of galaxy groups with masses less than 10$^{14}$ \Mdoth show a
significant population of S0 galaxies and passive spiral galaxies
\citep[McGee et al., in prep]{wilmanS0}, which are hard to reconcile
with their model.  On the other hand, the lower value of
\Mtrunc$\sim10^{13}$\Mdoth\ that we advocate might have trouble
explaining the large fraction of galaxies with [OII] emission in the
\citet{poggianti} clusters at $z\sim 0.5$.  Undoubtedly both models
are greatly oversimplified and, moreover, there are important
systematic uncertainties in the current data (especially in
determining cluster masses and galaxy star formation rates) and
statistical limitations resulting from small sample sizes.

Similar constraints have also been derived in the past from
observations of radial gradients in clusters. \citet{balogh_model}
used n-body simulations of the infall of substructure into clusters
and concluded that, to match the radial gradients of star formation
rates, the star formation rates in cluster galaxies must decline on
the timescale of a few Gyrs after entering the cluster. Significantly,
they also found that the best match to radial gradients was provided
if the star formation rate in the galaxy began to decline as soon as
it was found in a dark matter structure of group-size or larger.
\citet{ellingson} took this a step further and investigated the
evolution of such gradients.  They determined that 'field-like'
galaxies became early type galaxies on a 2-3 Gyr timescale.
\citeauthor{ellingson} also inferred that if galaxies were transformed
on the 3 Gyr timescale, than the galaxy infall rate into clusters
between z $\sim$ 1.5 and z $\sim$ 0.5 must have declined by $\sim$ 20
$\%$. Our results suggest that the infall rate of galaxies into
clusters over the same span fell by $\sim$ 15 $\%$. This is a
surprisingly good agreement given the large observational
uncertainties at each step in this analysis.

Finally, it is instructive to reexamine the results of B09 in the
context of our results. We have previously shown that B09 finds a
lower fraction of galaxies within groups and clusters than we do
(\textsection \ref{sec-sample}). This is likely due to a subhalo
completeness level which varies as a function of environment.  They
use the global number density of subhalos above their mass threshold
and compare it against SDSS number densities to conclude that their
global magnitude limit is $\sim$ 0.3 L$_*$. However, when they compare
number densities of their subhalos within clusters with cluster
observations, they find that their cluster magnitude limit is $\sim$
0.5 L$_*$. Using the red galaxy luminosity function derived from a
large sample of galaxy clusters by \citet{Lu}, a magnitude cut of 0.5
L$_*$ instead of 0.3 L$^*$ reduces the number of cluster galaxies by
$\sim$ 40 $\%$. In other words, groups falling into their clusters
could have $\sim$ 40 $\%$ fewer galaxies than would be expected from a
consistent luminosity cut.  Indeed, we find that this is on the order
of the discrepancy between our results and those of B09. For instance,
we have shown that, for 10$^{14.2}$ \Mdoth\ clusters, $\sim$ 35 $\%$
of galaxies have been accreted through 10$^{13}$ \Mdoth\ halos at z=0,
while B09 find only 24\%.  While this disagreement is significant for
evaluating the role of preprocessing in cluster assembly, a bigger
factor is that the B09 clusters are not very massive. Indeed, these
are smaller than the bulk of well studied clusters at intermediate and
high redshift. We have extended their analysis to more massive
clusters and find, as B09 themselves anticipated, that group
pre-processing is potentially much more important for more massive clusters. 

\subsection{Towards a physically motivated model} \label{sec_sem}

We have shown, by following the accretion of galaxies into groups and
clusters, and making simple assumptions about the nature of
environmental effects on galaxies, that the halo mass at which
environmental effects begin to be induced on galaxies is approximately
10$^{12}$ -- 10 $^{13}$ \Mdoth, and the time those effects take to
manifest themselves is quite long ($>$ 2 Gyr).  Here, we address some
of the more important simplifications we have made in constructing
this model.

The first simplification is that we have assumed that an environmental
effect will have a unique signature on the properties of
galaxies. However, in comparing our model to, for instance, the
fraction of red galaxies in clusters, we must acknowledge that there
is more than one process which can make a galaxy red. In the local
Universe, observations suggest that nearly all galaxies with stellar
masses above 10$^{10}$ \Mdoth\ are red regardless of their environment
\citep{baldry06}.  However, as shown in Figure \ref{gal_redstel}, in
our simple model the most massive cluster galaxies would still be red,
a consequence of the fact that they have resided within massive dark
haloes for a long time. This, combined with the fact that the more
numerous low mass galaxies dominate the fraction of galaxies in a
cluster, indicate that this is not a large complicating factor.

\begin{figure}
\leavevmode \epsfysize=8cm \epsfbox{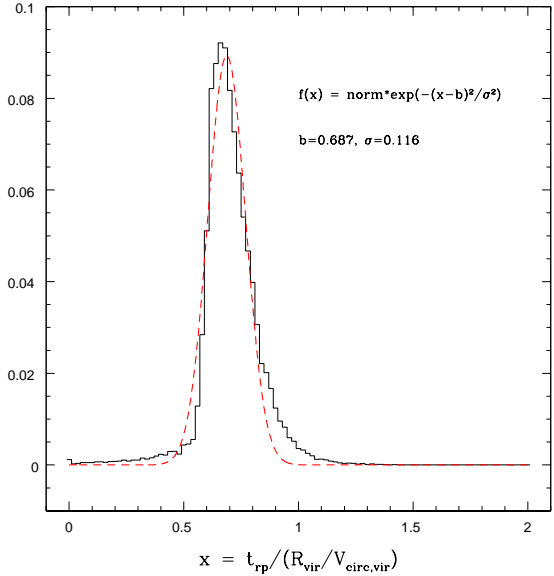}
\caption{The distribution of times, t$_{rp}$ for an infalling dark
  matter substructure to reach its pericentre from the virial radius,
  R$_{vir}$, of a halo with circular velocity V$_{circ,vir}$. The
  black line is the distribution of 10$^5$ randomly sampled orbits
  from \citet{Bensonorbit} and the dotted red line is the best fit
  gaussian.}
\label{orbits}
\end{figure}

Secondly, we have assumed that all galaxies display environmental
effects after a specific time \Ttrunc, regardless of their incoming
orbit. However, \citet{mccarthy} has shown in simulations that the
environmental effect on an infalling galaxy is dependent on the orbit
of that galaxy. \citeauthor{mccarthy} also showed that the bulk of the
environmental effect on an infalling galaxy occurs when the satellite
is at its pericentre. The size of this effect can be quantified by the
variation in the time it takes a galaxy to fall from the virial radius
to the pericentre of its orbit.  In Figure \ref{orbits}, we show the
distribution of times for a realistic distribution of infalling dark
matter substructure from \citet{Bensonorbit}, randomly sampled 10,000
times. The distribution is shown as a mass-independent quantity, along
with the best fit gaussian. A cluster of 10$^{14}$ \Mdoth\ has a
R$_{vir}$= 1.26 \Mpch\ and V$_{\rm circ}$=400 km/s, which translates
to a quite narrow distribution, with a dispersion of only $\sim$ 0.2
Gyrs. This will not have significant implications for a timescale
which is greater than 2 Gyrs. It is worth noting that the simulations
of \citeauthor{Bensonorbit} were not adequate to quantify the effect
of any host halo mass dependence of the orbital distribution, but the
indications are that this will not have a significant impact for our
purposes. Additionally, we have assumed that all galaxies entering a
massive halo feel similar environmental effects, however, galaxies
with large pericentric distances may not feel strong effects, and thus
predictions for the red fraction scatter and its mass dependence will
still benefit from proper tracing of orbits in the future.

Thirdly, we have assumed that the mass of a halo is the important
quantity driving any environmental effects. In fact, most of the
physical processes which could produce environmental effects are
likely more sensitive to X-ray gas density or temperature ($T_x$). For
galaxy clusters and massive groups the scatter in the M--$T_x$
relation is actually quite small ($\pm$ 30 $\%$ at $M = 10^{14.5}$
\Mdoth). However, the Mass -- X-ray Luminosity (M -- L) relation,
which is more sensitive to the gas density, does show significant
scatter at cluster mass scales
\citep{mccarthy_scatter,balogh_scatter}. But this scatter is driven by
properties of the group and cluster cores, while at the radius of a
typical galaxy pericentre (0.2-0.3 times the virial radius), the
scatter from system to system is quite small \citep{mccarthy_large,
sun_large}. So the bulk of ram pressure stripping will occur at radii
where the gas density has little scatter from system to
system. However, this analysis is limited to fairly massive groups and
clsuters, as measurement of X-ray properties for typical 10$^{12}$
\Mdoth\ haloes is quite difficult \citep{reiprich}. One theoretical
indication of the size of this effect in low-mass groups comes from
the scatter in the virial mass -- circular velocity relation, which is
approximately $\pm$ 15 $\%$ at M = 10$^{12}$ \Mdoth
\citep{bullock}. The circular velocity is more indicative of the depth
of the dark matter potential, and thus is likely more closely
correlated with the gas density.  Despite this, the size of this
scatter is likely not a huge source of uncertainty in our model, given
that we only make broad statements about the characteristic halo mass
scale. Given all of these results, it is encouraging that our model
does not appear too simple to give important insights to the behavior
of environmental effects.

The next step is to put this ad hoc model on a more physical basis. In
particular, in our model we have specified that galaxies within a host
halo are equally affected by environmental processes regardless of
their position within the halo. But because the cooling rate of gas in
a halo is density dependent, semi-analytic models treat galaxies at
the center of halos (centrals) different from those not in the center
(satellites). While this distinction is still a simplification
\citep{simha}, we point out the similarity between a model with a
\Mtrunc\ of 10$^{12}$ \Mdoth\ and one where the environmental effect
begins to occur when a galaxy becomes a satellite, as shown in Figure
\ref{predict_scatter} . Our most favored \Mtrunc\ model is essentially
equivalent to choosing a physically motivated central/satellite model.

We have also employed a fixed timescale for environmental effects to
occur. Ideally, we would like to link this timescale to a physically
motivated quantity, such as the orbital timescale of a galaxy in a
cluster or group. This mean timescale is approximately constant for
the groups and clusters in our mass range at a given redshift
epoch. However, because of the decreasing universe density with time,
at high redshift the orbital timescale is actually smaller by a factor
of $\sim$ (1+z)$^{3/2}$.  A timescale based on this would suggest that
at z=1.5 the timescale is $\sim$ 4 times shorter than the timescale at
z=0. Unfortunately, directly implementing a timescale based on the
orbital timescale would ignore several other complicating factors such
as the evolution of cluster gas density profiles and the evolution of
galaxy sizes and densities. Exploring these issues in a full
semi-analytic galaxy model is the important next step forward.

\section{Discussion and Conclusions} \label{sec-conclude}

We have used the stellar mass and merger trees produced by the
semi-analytic galaxy catalogues of F08 to follow the accretion of
galaxies into groups and clusters at four different redshift epochs
(z=0,0.5,1.0 and 1.5) for samples of galaxies with stellar mass
$M>10^9$\Mdoth. By tracking galaxies through the hierarchy of
structure formation we are able to examine the effect that
environmental processes may have on the galaxy population of groups
and clusters. Further, by adopting a simple model for the
environmental effects, we are able to make strong claims about the
timescale and mass threshold on which environmental effects occur.
Our main results are summarized as follows:

\begin{itemize}
\item Clusters at all redshifts examined have had a significant
  fraction of their galaxies accreted through galaxy groups. For
  instance, 10$^{14.5}$ \Mdoth\ mass clusters at z=0 have had $\sim$
  40$\%$ of their galaxies (M$_{\rm stellar} > 10^{9}$ \Mdoth )
  accreted through halos with masses greater than 10$^{13}$ \Mdoth. At
  higher redshifts fewer galaxies are accreted through massive
  halos. Only $\sim$ 25 $\%$ of galaxies have been accreted through
  10$^{13}$ \Mdoth\ into 10$^{14.5}$ \Mdoth\ mass clusters at z=1.5.

\item We find only a moderate difference in the stellar mass accretion
  history and the galaxy accretion history at high cluster mass. That
  is, more massive galaxies are accreted preferentially through
  groups. While 45$\%$ of galaxies in 10$^{15}$ \Mdoth\ mass clusters
  at z=0 are accreted through halos with masses greater than 10$^{13}$
  \Mdoth, 50$\%$ of the stellar mass is accreted through the same halo
  mass range. Contrary to the study of \citet{berrier}, we do not see
  a large difference between the galaxy assembly of clusters and the
  mass assembly of clusters.
 
\item Following from the previous point, we find that the extent to
  which galaxies are pre-processed in groups before falling into
  clusters depends on the stellar mass of the infalling galaxy. For a
  10$^{14.5}$ \Mdoth\ mass cluster, 73 $\%$ of galaxies with stellar
  masses greater than 10$^{10.5}$ \Mdoth\ are accreted through
  10$^{12}$ \Mdoth\ systems, while only 50$\%$ of 10$^{9}$ to
  10$^{10}$ \Mdoth\ are accreted through the same systems. Further, we
  find that in the accretion through group sized halos increases at
  late times when compared to the accretion into the cluster during
  early times.

\item We have shown that the fraction of isolated galaxies infalling
  into z=0 groups and clusters is remarkably independent of the final
  cluster mass. 5-6 $\%$ of the final cluster galaxies are accreted
  per Gyr for the last 10 Gyrs. Thus if a galaxy begins to be affected
  by its environment soon after becoming a satellite galaxy, and the
  time it takes for that effect to manifest itself is constant with
  halo mass, then a similar fraction of galaxies are affected in each
  cluster above a halo mass of 10$^{13}$ \Mdoth.

\item Despite the previous result, observing a cluster of the same
  halo mass at each redshift epoch implies different accretion rates
  of isolated galaxies, from 5-6 $\%$ per Gyr at z=0 to 15$\%$ per Gyr
  at z=1.5. Thus, in effect, the Butcher Oemler effect may be
  qualitatively explained by the shorter time available for cluster
  assembly at higher redshift.

\item We find that combining the simple observations of the existence
  of a significant Butcher Oemler effect at z=0.5 and the observations
  that galaxies within groups display significant environmental
  effects with galaxy accretion histories justifies striking
  conclusions. Namely, that the dominant environmental process must
  begin to occur in halos of 10$^{12}$ -- 10$^{13}$ \Mdoth\ and act
  over timescales of $>$ 2 Gyrs. This supports a long lifetime, gentle
  mechanism like strangulation.

\item This simple model predicts that by z=1.5 galaxy groups and
  clusters will display little to no environmental effects. This
  conclusion may have limit the effectiveness of red sequence cluster
  finding methods at high redshift.

\end{itemize}

In essence, we have seen that systematic observations of intermediate
and high redshift clusters and groups have the power to strongly
constrain the mechanisms which induce environmental transformations on
galaxies. However, because of the significant cluster to cluster
variations in environmental effects, it is important that the method
for selecting galaxy clusters and groups for observation must be
easily and accurately reproducible in cosmological simulations. Only
this will allow the careful testing of models against observations. In
a future paper we will compare the best available data on groups and
clusters at a variety of redshift epochs to further constrain the
dominant environmental processes.
 
Significant progress on the implications of strangulation and the
physical processes involved will need more extensive hydrodynamical
simulations.  The simulations of ram pressure stripping of the hot
haloes of infalling galaxies by \citet{mccarthy} is a significant step
forward. However, there are important unknowns. In particular, how
effective are low mass group halos in stripping the infalling
galaxies? Unfortunately, this is sensitively dependent on how the gas
is distributed in both the infalling galaxy and the group
halos. Indeed, the effectiveness of strangulation is also dependent on
the strength of star formation feedback, and how reheated galaxy gas
is distributed and stripped from the galaxy. The behavior of galaxies
within small groups which subsequently fall into massive clusters is
also unclear. To what extent are galaxies ``shielded'' by their local
group from further gas stripping? Encouragingly, large scale
hydrodynamical simulations are beginning to be able to address some of
these questions \citep[e.g.][]{crain}.

So, while there is much room for improvement in understanding the
details of galaxy -- environment interactions, our results have shown
that the galaxy accretion histories of groups and clusters combined
with a simple model strongly suggest that the dominant environmental
effect occurs over long time scales and is effective in low mass
halos. In a future paper, we will examine these insights by making a
quantitative comparison between semi-analytic models and the best
available cluster, group and field data to z $\sim$ 1.

\section*{Acknowledgments}

We thank the GALFORM team for allowing access to the semi-analytic
galaxy catalogues used in this paper and the Virgo Collaboration for
carrying out the Millennium simulation. We thank the referee for
comments which improved the paper and we thank Erica Ellingson and Mike
Hudson for discussions at an early stage of this work. MLB
acknowledges support from an NSERC Discovery Grant. IGM acknowledges
support from a Kavli Institute Fellowship.

\bibliography{ms}

\section*{Appendix: Full accretion histories}

\begin{figure*}
\includegraphics[width=\textwidth]{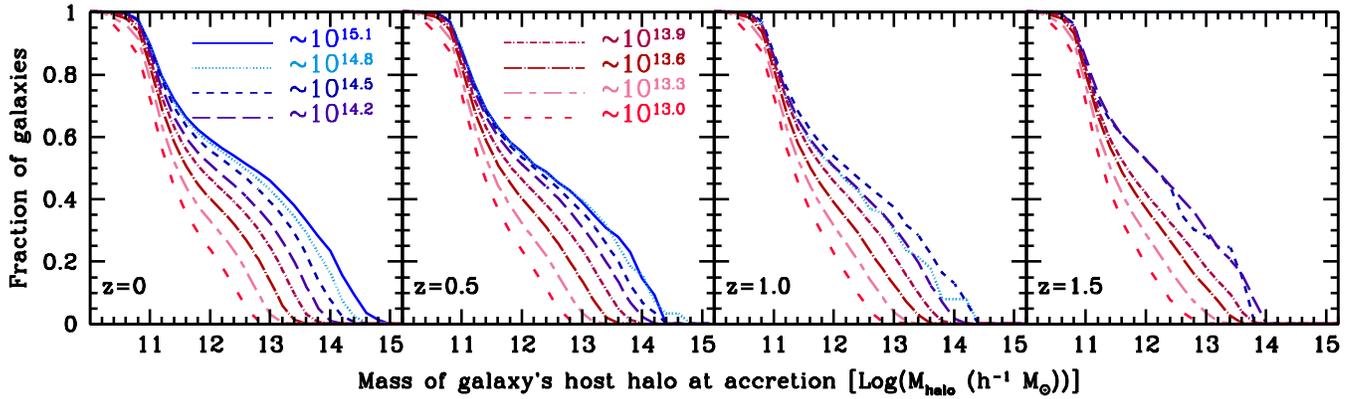}

\caption{The cumulative distribution of cluster galaxies which reside
  in a host halo of a given mass at the time of accretion into the
  final cluster halo. In the left panel is the accretion history of 8
  composite clusters of a given final host mass at z=0, while in the
  left middle (right middle) [right] panel is a separate final cluster
  sample at z=0.5 (z=1) [z=1.5]. All cluster galaxies have final
  stellar masses of M $>$ 10$^{9} M_\odot$. The mass range bins were
  defined in Table 1, and are shown for all bins containing more than
  one cluster.}
\label{gal_accr}
\end{figure*}

\begin{figure*}
\includegraphics[width=\textwidth]{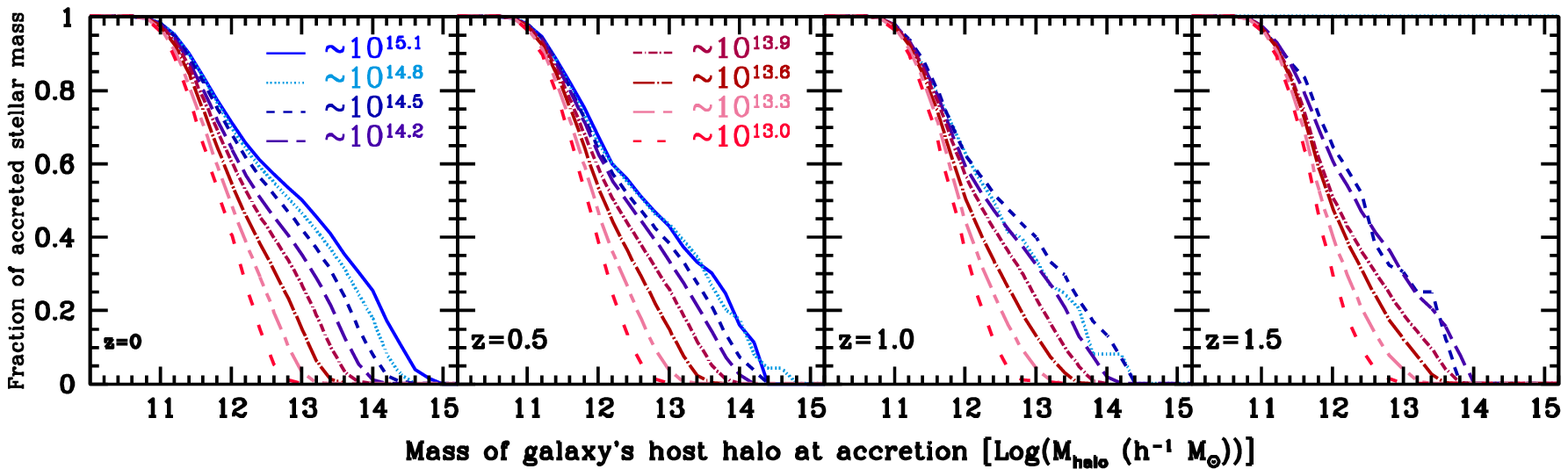}
\caption{The cumulative distribution of accreted stellar mass which
  reside in a host halo of a given size at the time of accretion into
  the final cluster halo. In the left panel, is the accretion history
  of 8 composite clusters of a given final host mass at z=0. The left
  middle (right middle ) [right] panel is for a separate final cluster
  sample at z=0.5 (z=1) [z=1.5]. All cluster galaxies have final
  stellar masses of M $>$ 10$^{9} M_\odot$. The mass range bins were
  defined in Table 1, and are shown for all bins containing more than
  one cluster.}
\label{stel_accr}
\end{figure*}

Here we show the complete accretion histories for each bin of cluster
mass and for all four redshift epochs. They are presented for both the
galaxy accretion (Figure \ref{gal_accr}) and for stellar mass
accretion (Figure \ref{stel_accr}). Figure \ref{gal_accr} shows the
cumulative distribution of accreted cluster galaxies which reside in a
host halo mass of a given size prior to accretion into the final
cluster at each of four epochs of observation. Because galaxies are on
average more massive in more massive halos, this accretion history
does not agree completely with dark matter accretion
histories. Therefore, we present the complete stellar mass accretion
histories in Figure \ref{stel_accr}. Again, this shows the cumulative
distribution of the accreted stellar mass as a function of the
galaxy's host halo mass at the time of accretion.

\end{document}